\documentclass[9pt]{extarticle}
\usepackage[T1]{fontenc}

\usepackage{spconf}
\usepackage{amsmath,graphicx,tikz,amssymb,amsfonts,bm,mathtools,mathabx}
\usepackage{multicol}
\usepackage{amsfonts}
\usepackage{amssymb}
\usepackage{amsthm}
\usepackage{float}
\usepackage{comment}
\usepackage[linkcolor=mar,colorlinks=true]{hyperref}
\hypersetup{colorlinks,breaklinks,
            linkcolor=[rgb]{0.5,0,0},
            citecolor=[rgb]{0.39,0.7,1.0}}
\usepackage{cleveref}
\usepackage{cite}
\usepackage{enumitem}
\usepackage{outlines}
\usepackage[ruled,vlined,linesnumbered]{algorithm2e}
\usepackage{algpseudocode}
\usepackage[skip=0pt]{caption}
\usepackage[skip=0pt]{subcaption}
\usepackage{bm}
\newtheorem{theorem}{Theorem}

\newtheoremstyle{reduced}
  {6pt} 
  {6pt} 
  {} 
  {} 
  {\bfseries} 
  {.} 
  {.5em} 
  {} 
\newtheorem{lemma}[theorem]{Lemma}

\theoremstyle{reduced}
\newtheorem{assumption}{Assumption}
\newtheorem{remark}{Remark}

\newcommand*{\trans}{{\mathsf{T}}}
\newcommand*{\hermtr}{{\mathsf{H}}}



\renewenvironment{thebibliography}[1]{%
   \begin{oldthebibliography}{#1}%
     \setlength{\itemsep}{0ex}%
}%
{%
   \end{oldthebibliography}%
}

\title{Model-Free Learning of Optimal Beamformers for \\
Passive IRS-Assisted Sumrate Maximization
\vspace{-10pt}}
\name{Hassaan Hashmi, Spyridon Pougkakiotis, Dionysios S. Kalogerias
\vspace{-10pt}}
\address{Department of EE, Yale University, New Haven, USA\\
 \small{\tt \{\href{mailto:hassaan.hashmi@yale.edu}{hassaan.hashmi}, \href{mailto:spyridon.pougkakiotis@yale.edu}{spyridon.pougkakiotis}, \href{mailto:dionysis.kalogerias@yale.edu}{dionysis.kalogerias}\}@yale.edu}
\vspace{-14pt} }

\begin{document}
\setlist[itemize]{noitemsep, topsep=1pt}
\setlist[itemize]{leftmargin=*}
\maketitle
\begin{abstract}
\vspace{-4pt}
Although Intelligent Reflective Surfaces (IRSs) are a cost-effective technology promising high spectral efficiency in future wireless networks, obtaining optimal IRS beamformers is a challenging problem with several practical limitations. Assuming fully-passive, sensing-free IRS operation, we introduce a new data-driven \textit{Zeroth-order Stochastic Gradient Ascent (ZoSGA)} algorithm for sumrate optimization in an IRS-aided downlink setting. ZoSGA does not require access to channel model or network structure information, and enables learning of optimal long-term IRS beamformers jointly with standard short-term precoding, based only on conventional \textit{effective} channel state information. Supported by state-of-the-art (SOTA) convergence analysis, detailed simulations confirm that ZoSGA exhibits SOTA empirical behavior as well, consistently outperforming standard fully model-based baselines, in a variety of scenarios. 
\end{abstract}
\vspace{-5pt}

\vspace{-10pt}
\section{Introduction}\label{sec: intro}
\begin{figure}[!t]
  \centering
  \centerline{\includegraphics[width=1.92in]{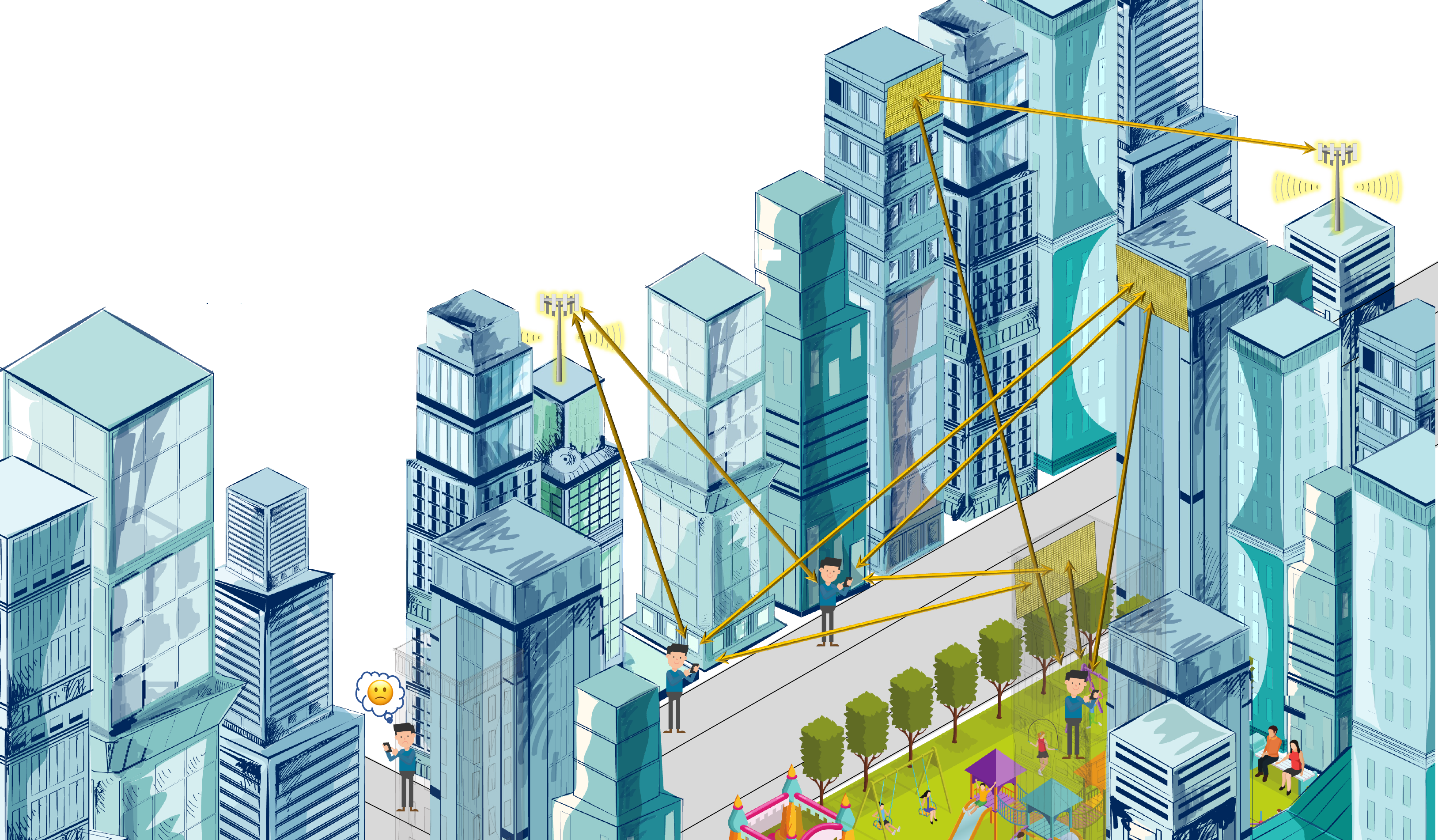}}
\vspace{4bp}
\caption{Realization/Concept of an IRS-aided Wireless Network.}
\label{fig:sketch}
\end{figure}

\vspace{-8pt}
Intelligent Reflective (or Reconfigurable Intelligent) Surfaces (IRSs or RISs) are planar surfaces comprised of passive reflective elements with tunable phase-shifts and amplitudes. Seen as beamforming arrays, IRSs present a promising solution to mitigate sharp drops in Quality-of-Service (QoS) in the absence of line of sight for highly directional mmWave signals, and beyond. Such a typical scenario is visualized in Fig.\:\ref{fig:sketch}, where users linked with an Access Point (AP) are enjoying improved QoS due to the IRSs assisting the network.

In IRS-aided communications, the goal is to \textit{optimally tune} the IRS elements along with other possible resources (such as AP precoders) to optimize a certain system utility. A standard case is that of the \textit{weighted sumrate utility} in a MISO downlink scenario (see Fig.\:\ref{fig:sketch}), where the goal is to maximize the total downlink rate of a number of users (i.e., terminals) actively serviced by an AP, while passively aided by one or multiple IRSs.
Then, the objective is to jointly optimize IRS amplitudes/phase-shifts and AP precoders, subject to certain power constraints.
While AP precoders are usually continuous-valued, IRS phase-shifts can be either quantized \cite{kamoda2011pin_diodes} or continuously varying \cite{zhao2013varactor1, araghi2022varactor2}. In this paper, we focus on the latter, i.e., on optimizing for continuous IRS phase-shift variations, however strictly assuming fully-passive IRS elements with no sensing capabilities or extra hardware or scheduling requirements.

Recently, various deep learning driven methods have been proposed for IRS-aided beamforming optimization. Offline learning approaches have been proposed primarily to estimate the Channel State Information (CSI) using labeled datasets with Function Approximators (FAs) \cite{offline:taha2019,offline:balevi2020,offline:yang2021,offline:zhang2021}. In contrast, Deep Reinforcement Learning (DRL) methods have been used for joint beamforming optimization, either with quantized \cite{dqn:mismar2019,dqn:taha2020, dqn:yang2020} or continuous \cite{ddpg:huang2020, ddpg:yang2020} actions. Though such \textit{end-to-end} DRL methods do not require intermediate CSI estimates or active IRS sensing, they do require FAs to at least approximate value functions, jointly modeling the whole optimization task, including beamforming controllers. On the one hand, choosing such FAs without explicit domain knowledge usually results in increased problem complexity, non-interpretability and lack of robustness, whereas incorporating specific domain knowledge often results in overfitting, hindering versatility and transferability to distinct environments. Further, as such approaches primarily consider reactive IRS tuning (i.e., phase shifts depend on observed CSI), they naturally incur increased power consumption for perpetual IRS control.

To improve complexity, efficiency and robustness, \textit{model-based two-timescale} 
approaches 
have also been proposed, in which reactive (AP) precoding is combined with static (i.e., non-reactive) IRS beamforming \cite{guo2020larsson,zhao2020tts, zhao2021qos, yang2021sca, liu2021unfold}. Although potentially effective, such methods rely on optimizing problem surrogates and leverage detailed knowledge of channel models and spatial network configurations, often requiring active sensing at the IRSs. Thus, they are not applicable to different settings without complete remodeling of the environment.

In this paper, 
we develop a new \textit{Zeroth-order Stochastic Gradient Ascent (ZoSGA)} two-timescale algorithm for weighted sumrate optimization in the MISO downlink setting through \textit{purely data-driven learning} of optimal IRS beamformers, and joint reactive AP precoding via the standard WMMSE algorithm \cite{wmmseShi2011}. 
ZoSGA follows the two-timescale paradigm while being \textit{completely FA- and model-free}, and relies on minimal zeroth-order reinforcement (i.e., system probing), as well as user-experienced effective CSI (see Section \ref{ProblemFormulation}), conventionally available in mutli-user downlink beamfoming, regardless of the number or spatial configuration of the IRSs.
In fact, ZoSGA allows treating the IRSs assisting the network as fully passive (though tunable) reflecting elements, and applies to different networking scenarios seamlessly. 

We analyze ZoSGA under mild regularity conditions and establish a state-of-the-art (SOTA) convergence rate of order $\mathcal{O}(\sqrt{S}\epsilon^{-4})$, where $S$ is the number of IRS tunable parameters and $\epsilon$ is the suboptimality target. Then, we numerically demonstrate that ZoSGA also achieves SOTA empirical performance in a variety of scenarios, substantially outperforming the two-timescale method of \cite{zhao2020tts}, which is a standard \textit{model-based} baseline for the problem under study, assuming full knowledge of the underlying channel model and spatial network configuration. Overall, ZoSGA sets a new SOTA for beamforming optimization in IRS-aided communications.

\textit{Note:} Proofs are omitted due to lack of space and will be included in a journal paper currently under preparation. Fully reproducible code is available at \textcolor{cyan}{github.com/hassaanhashmi/zosga-irs}.

\clearpage
\vspace{-10pt}
\section{Problem Formulation}\label{ProblemFormulation}
\vspace{-8pt}
Capitalizing on the standardized IRS-Aided communication setting depicted in Fig.\:\ref{fig:sketch}, our goal is to maximize the total downlink rate of $K$ users actively serviced by an AP with $M$ antennas, while passively aided by one or multiple IRSs, arbitrarily spatially placed. As explained in Section \ref{sec: intro}, we assume dynamic (i.e., reactive) AP beamformers (also called precoders), while the IRS beamformers are realistically viewed as static (i.e., non-reactive) elements, tunable by a parameter vector $\bm{\theta}\in \mathbb{R}^S$, encompassing whatever propagation feature is learnable on the IRSs present in the network. For instance, $\bm{\theta}$ might refer to amplitudes, phases, or both, or even realistic \textit{physical elements} of an IRS, such as tunable varactor capacitance elements, see, e.g., \cite{zhao2013varactor1, araghi2022varactor2}.
We strictly make no sensing  assumptions on the IRSs, i.e., the IRSs are completely passive network elements.

Each user $k=1,\ldots,K$ experiences a random \textit{effective channel} denoted by $\mathbf{h}_k\left(\bm{\theta},\omega \right)$, indexed by the IRS parameter vector $\bm{\theta}$ as well as a \textit{state of nature} $\omega\in \Omega$ describing \textit{unobservable} random propagation patterns for each value of $\bm{\theta}$. In other words, $\mathbf{h}_k\left(\bm{\theta},\omega \right)$ is a random channel field with spatial variable $\bm{\theta}$. We make the standard assumption that the effective channels $\mathbf{h}_k\left(\bm{\theta},\omega \right),k=1,\ldots,K$ are known to the AP at the time of transmission \cite{guo2020larsson,zhao2020tts}. Note that the implementation complexity of estimating effective channels in our setting is exactly the same as that in conventional multi-user downlink beamforming (i.e., involving no IRSs), \textit{regardless} of the number and/or spatial configuration of the IRSs; no extra hardware or customized scheduling schemes (such as those in \cite{zhao2020tts}) are required on either the AP or the IRSs assisting the network.





\par The QoS of user $k$ is measured by the corresponding SINR, i.e.,
\begin{align}\nonumber
\begin{aligned} \label{eq:2}
\text{SINR}_k(\mathbf{W},\mathbf{h}_k(\bm{\theta}, \omega )) \triangleq \frac{\left|\mathbf{h}_k^\hermtr(\bm{\theta}, \omega)\mathbf{w}_k\right|^2}{{\sum}_{{j \in \mathbb{N}_{K}^+ \setminus k}}\left|\mathbf{h}^\hermtr_k(\bm{\theta}, \omega)\mathbf{w}_j\right|^2 + \sigma^2_k},
\end{aligned}
\end{align}
where $\mathbf{W} {=} \begingroup \setlength\arraycolsep{2pt} \begin{bmatrix} \mathbf{w}_1 & \mathbf{w}_2 & \cdots & \mathbf{w}_K \end{bmatrix} \endgroup \in \mathbb{C}^{M \times K}$, $\mathbf{w}_k$ is a transmit precoding vector and $\sigma^2_k$ is the noise variance for user $k$, respectively. Then, the \textit{weighted sumrate utility} of the network is defined as
\begin{equation}\nonumber
{F}(\mathbf{W},\mathbf{H}(\bm{\theta},\omega)) \triangleq \sum_{k=1}^K \alpha_k \log_2\left(1 + \textnormal{SINR}_k\left(\mathbf{w},\mathbf{h}_k\left(\bm{\theta},\omega  \right)\right)\right),
\end{equation}
\noindent with $\mathbf{H} = \begin{bmatrix} \mathbf{h}_1 \ldots \mathbf{h}_K \end{bmatrix}\in \mathbb{C}^{M\times K}$, and $\alpha_k\ge0$ the weight associated with user $k$.
We are interested in maximizing the sumrate of the network jointly by selecting instantaneous-optimal dynamic AP precoders $\mathbf{W}$, and on-average-optimal static IRS beamformers $\bm{\theta}$ \cite{guo2020larsson,zhao2020tts}, i.e., we are interested in the problem
\begin{equation}\label{eq:1}
\boxed{\max_{\bm{\theta} \in \mathcal{K}} \mathbb{E} 
\bigg\{ 
\max_{\mathbf{W}: \|\mathbf{W}\|_F^2 \leq P} 
{F}\left(\mathbf{W},\mathbf{H}(\bm{\theta},\omega)\right)
\bigg\},}
\end{equation}
where $\|\cdot\|_F^2$ denotes the Frobenius norm, $P>0$ is a total power budget at the AP, and $\mathcal{K}$ is a compact feasible set. Problem \eqref{eq:1} is an instance of a  \textit{$2$-stage stochastic program} \cite{SIAM:Shapiro_etal}. Accordingly, we may interpret optimization over $\bm{\theta}$ as the first-stage problem, where the decision-maker optimizes the IRS parameters on-average and \textit{before} the actual effective channels are revealed, whereas optimization over $\mathbf{W}$ constitutes the second stage problem which is solved \textit{after} randomness is revealed to the decision-maker, therefore interpreting optimal AP beamforming as (optimal) \textit{recourse actions}.

\begin{figure*}[!ht]
\vspace*{-16pt}
\begin{align}\tag{D}
\begin{aligned} \label{eq:big} 
     \bm{D}(\mathbf{W},\mathbf{H}(\bm{\theta}, \omega))
    = \sum_{k = 1}^K\alpha_k\frac{\mathbf{z}^\hermtr\left[\left(\sum_{j \in \mathbb{N}_{K}^+ \setminus k}\left|\mathbf{z}^\hermtr\mathbf{w}_j\right|^2+\sigma^2_k\right)\mathbf{w}_k{\mathbf{w}_k}^\hermtr - \left|\mathbf{z}^\hermtr\mathbf{w}_k\right|^2 \sum_{j \in \mathbb{N}_{K}^+ \setminus k} \mathbf{w}_j{\mathbf{w}_j}^\hermtr \right]}{\ln(2)(\sum_{j \in \mathbb{N}_{K}^+ \setminus k}|\mathbf{z}^\hermtr\mathbf{w}_j|^2+\sigma^2_k)^2 + |\mathbf{z}^\hermtr\mathbf{w}_k|^2(\sum_{j \in \mathbb{N}_{K}^+ \setminus k}|\mathbf{z}^\hermtr\mathbf{w}_j|^2+\sigma^2_k)}\Bigg\vert_{\mathbf{z}=\mathbf{H}(\bm{\theta}, \omega)}.
\end{aligned}
\end{align}
\hrulefill
\vspace*{4pt}
\end{figure*}

\begin{figure*}[!ht]
\vspace*{-16pt}
\begin{align} \tag{G}
\begin{aligned} \label{eq:big2}
  \dfrac{\mathbf{G}_{\mu}(\bm{\theta},\omega,\mathbf{u})}{2} \triangleq &
   \left(\frac{\Re\left(\mathbf{\Delta}_{\mu}(\bm{\theta},\omega,\mathbf{u})\right)}{2\mu} \mathbf{u}^\trans \right)^\trans\left(\Re\left( \bm{D}(\mathbf{W}^*,\mathbf{H}(\bm{\theta},\omega)) \right)\right)^\trans  + \left(\frac{\Im\left(\mathbf{\Delta}_{\mu}(\bm{\theta},\omega,\mathbf{u})\right)}{2\mu} \mathbf{u}^\trans \right)^\trans\left(\Re\left(j \bm{D}(\mathbf{W}^*,\mathbf{H}(\bm{\theta},\omega)) \right)\right)^\trans.
\end{aligned}
\end{align}
\hrulefill
\vspace*{-8pt}
\end{figure*}
\vspace{-10pt}
\section{An inner-outer optimization method}
\vspace{-6pt}
\par We would like to devise a gradient-inspired method for  tackling problem \eqref{eq:1}, which circumvents the need for accessing first-order gradient information of the effective channel function $\mathbf{H}(\cdot,\omega)$; in principle, this is unknown. To that end, we derive an inner-outer scheme for the solution of the two-stage problem in \eqref{eq:1}. At each (outer) iteration (i.e., for each state of nature $\omega$), we employ an oracle to find an approximate solution to the inner problem, which is then utilized to derive a model-free gradient approximation for the outer problem using zeroth-order reinforcement (i.e., probing) on the involved effective channel. This approach completely bypasses the need of a model for the effective channel, and offers great flexibility as to how the inner problem can be approximately solved.

\subsection{Tackling the Inner Problem (AP Precoding)} \label{subsec: inner opt}
\par 
To obtain an approximate solution for the (nonconvex) inner maximization problem, we heuristically employ the well-known weighted minimum mean squared error (WMMSE) algorithm \cite{wmmseShi2011},
simultaneously implementing AP precoding. For technical purposes, we also introduce the following assumption.
\begin{assumption} \label{assumption: inner problem solution}
Given any $\bm{\theta} \in \mathcal{K}$ and almost every (a.e.) $\omega \in \Omega$, we have access to an oracle that yields an optimal solution $\boldsymbol{W}^*(\boldsymbol{\theta},\omega) \in \arg\max_{\mathbf{W}\colon \|\mathbf{W}\|_F^2 \leq P} {F}(\mathbf{W},\mathbf{H}(\bm{\theta},\omega))$.
\end{assumption}
\noindent In practice Assumption \ref{assumption: inner problem solution} may be restrictive but, as verified in Section \ref{sec: Simulations}, the empirical effectiveness of the proposed algorithm is not hindered. Nevertheless, this assumption is crucial for the derivation of our algorithm and its subsequent theoretical analysis.

\subsection{Tackling the Outer Problem (IRS Beamforming)} \label{subsec: outer optimization}

\par Assuming the availability of an oracle providing an optimal solution to the inner problem at any $(\bm{\theta},\omega)$, say $\mathbf{W}^*(\bm{\theta},\omega)$ (in practice obtained approximately via WMMSE), 
we can write problem \eqref{eq:1} as 
\begin{equation} \label{eqn: outer problem}
    \begin{split}
        \max_{\bm{\theta} \in \mathcal{K}}&\ \mathbb{E}\left\{{F}\left(\mathbf{W}^*(\bm{\theta},\omega),\mathbf{H}(\bm{\theta},\omega)\right)\right\}.
    \end{split}
\end{equation}
\noindent 
As we would like to derive a gradient-ascent-like scheme for solving \eqref{eqn: outer problem}, let us make the following assumptions on the effective channel.
\begin{assumption} \label{assumption: channel properties}
For almost all $\omega$, the channel function $\mathbf{H}(\cdot,\omega)$ is uniformly bounded, twice continuously-differentiable, $L_{\mathbf{h},0}$-Lipschitz, with $L_{\mathbf{h},1}$-Lipschitz gradients.
\end{assumption}
\noindent We note that Assumption \ref{assumption: channel properties} imposes regularity conditions mainly required for the grounded development of our optimization scheme and for its convergence analysis (later in Section \ref{sec: conv anal}). Also, observe
that the boundedness assumption is natural, since (IRS-aided) wireless channels are always bounded in practice. While we usually have no information on the analytical properties of the effective channel, Assumption \ref{assumption: channel properties} is easily satisfied in widely used channel models of IRS-aided systems, see, e.g., \cite{zhao2020tts} as well as Section \ref{sec: Simulations}.
\par 
From Assumptions \ref{assumption: inner problem solution}--\ref{assumption: channel properties}, and the compactness of $\mathcal{K}$, we obtain
\begin{equation} \label{eqn: gradient of the compositional objective}
\begin{split}
    &\nabla_{\bm{\theta}} \mathbb{E}\left\{ {F}\left(\mathbf{W}^*(\bm{\theta},\omega),\mathbf{H}(\bm{\theta},\omega)\right)\right\} \\ &\quad = \mathbb{E}\left\{\nabla_{\bm{\theta}} {F}\left(\mathbf{W}^*(\bm{\theta},\omega),\mathbf{H}(\bm{\theta},\omega)\right) \right\}\\   &\quad =\mathbb{E}\left\{\nabla_{\bm{\theta}} {F}\left(\mathbf{W},\mathbf{H}\left(\bm{\theta},\omega\right)\right)\mid_{\mathbf{W} = \mathbf{W}^*(\bm{\theta},\omega)} \right\},
    \end{split}
\end{equation}
\noindent where we used \textnormal{\cite[Theorem 7.44]{SIAM:Shapiro_etal}}, and the \emph{implicit function theorem} (e.g., see \textnormal{\cite{Springer:DonRock}}). Next, we derive the gradient of ${F}(\mathbf{W},\mathbf{H}(\cdot,\omega))$.

\subsubsection{Gradient Representation for the Outer Problem}
\par We notice that ${F}(\mathbf{W},\cdot)$ takes a complex input and hence we need an appropriate generalization of the gradient for it. To that end, we utilize the so-called \emph{Wirtinger calculus} (see \cite{arXiv:Kreutz-Delgado}). In the following lemma we derive the full \textit{compositional} gradient of ${F}(\mathbf{W},\mathbf{H}(\bm{\theta},\omega))$ by following the developments in \cite[Section 4]{arXiv:Kreutz-Delgado}.
\begin{lemma} \label{lemma: gradient of compositional}
For every $\bm{\theta} \in \mathcal{K}$, $\mathbf{W} \in \mathbb{C}^{M\times K}$ and $\omega\in \Omega$, the gradient of ${F}\left(\mathbf{W},\mathbf{H}(\bm{\theta},\omega)\right)$ with respect to $\bm{\theta}$ reads
\begin{equation*}
\begin{split} \nabla_{\bm{\theta}} &{F}\left(\mathbf{W},\mathbf{H}(\bm{\theta},\omega)\right)\\
&\quad = 2\nabla_{\bm{\theta}} \Re\left(\mathbf{H}(\bm{\theta},\omega)\right)\left(\Re\left(\bm{D}\left(\mathbf{W},\mathbf{H}(\bm{\theta}, \omega)\right) \right)\right)^\trans \\ &\qquad + 2\nabla_{\bm{\theta}} \Im\left(\mathbf{H}(\bm{\theta},\omega)\right)\left(\Re\left(j\bm{D}\left(\mathbf{W},\mathbf{H}(\bm{\theta}, \omega)\right) \right)\right)^\trans,
\end{split}
\end{equation*}
\noindent where $\bm{D}(\mathbf{W},\mathbf{H}(\bm{\theta}, \omega))$ is given in \textnormal{\eqref{eq:big}}. Additionally, there exists a constant $B_F > 0$ such that $\left\|\bm{D}(\mathbf{W},\mathbf{H}(\bm{\theta}, \omega))\right\| \leq B_F$ for every $(\mathbf{W},\bm{\theta}) \in \mathbb{C}^{M\times K} \times \mathcal{K}$ and a.e. $\omega \in \Omega$, where $\mathbf{W}$ is feasible for \eqref{eq:1}.
\end{lemma}
\subsubsection{Model-Free Jacobian Approximation of the Effective Channel}
\par Assuming the lack of availability of any first-order information of $\mathbf{H}(\cdot,\omega)$ for any $\omega \in \Omega$, we will employ a zeroth-order scheme in order to obtain a Jacobian estimate of $\nabla_{\bm{\theta}} \mathbf{H}(\bm{\theta},\omega)$, using which we can solve \eqref{eqn: outer problem} via a stochastic gradient ascent scheme. The proposed method will be based on Jacobian estimates arising from a two-point stochastic evaluation of $\mathbf{H}(\cdot,\omega)$ (similar to, among others, \cite{SIAMOPT:KalogeriasPowellZerothOrder,CompMath:Nesterov_etal,arXiv:Pougk-Kal}). From Assumption \ref{assumption: channel properties}, we can first write
\[\nabla_{\bm{\theta}} \mathbf{H}(\bm{\theta},\omega) = \nabla_{\bm{\theta}} \Re\left(\mathbf{H}(\bm{\theta},\omega)\right) + j \nabla_{\bm{\theta}} \Im\left(\mathbf{H}(\bm{\theta},\omega)\right).\]
\noindent We approximate this gradient using only function evaluations of $\mathbf{H}(\cdot,\omega)$ (i.e. by minimally probing the network). For each $\omega \in \Omega$, let $\mathbf{u} \sim \mathcal{N}\left(\mathbf{0}, \mathbf{I}\right)$. Given a smoothing parameter $\mu > 0$, we define
\begin{equation} \label{eqn: ZO gradient approximation}
\begin{split}
\hspace{-2pt}\nabla^\mu_{\bm{\theta}} \mathbf{H}(\bm{\theta},\omega) \triangleq&\ \mathbb{E}\left\{\frac{\mathbf{H}\left(\bm{\theta} + \mu \mathbf{u},\omega\right) - \mathbf{H}\left(\bm{\theta}-\mu \mathbf{u},\omega\right)}{2\mu} \mathbf{u}^\trans \right\}^\trans.
\end{split}
\end{equation}
\noindent  Using Lemma \ref{lemma: gradient of compositional}, we substitute the Jacobian of the effective channel via the zeroth-order approximation given in \eqref{eqn: ZO gradient approximation} to obtain the zeroth-order compositional quasi-gradient (an approximation of \eqref{eqn: outer problem}) as
\begin{equation*}
\begin{split} {\hat{\nabla}}_{\bm{\theta}}^{\mu} &{F}\left(\mathbf{W}^*,\mathbf{H}(\bm{\theta},\omega)\right)\\
&\quad = 2\nabla^\mu_{\bm{\theta}}\Re\left(\mathbf{H}(\bm{\theta},\omega)\right)\left(\Re\left(\bm{D}\left(\mathbf{W}^*,\mathbf{H}(\bm{\theta},\omega)\right) \right)\right)^\trans \\ &\qquad +  2\nabla^\mu_{\bm{\theta}}\Im\left(\mathbf{H}(\bm{\theta},\omega)\right)\left(\Re\left(j \bm{D}\left(\mathbf{W}^*,\mathbf{H}(\bm{\theta},\omega)\right)  \right)\right)^\trans.
\end{split}
\end{equation*}
\par We are now able to derive the proposed inner-outer scheme for the solution of problem \eqref{eq:1}. At every iteration $t$, we draw independent and identically distributed (i.i.d.) samples $\omega^{t+1}$ and $\mathbf{u}^{t+1}$, and compute an ``optimal" inner solution $\mathbf{W}^*(\bm{\theta}^{t},\omega^{t+1})$, as described in Section \ref{subsec: inner opt}. Then, by probing the system twice, we can compute a zeroth-order \textit{stochastic} gradient estimator as shown in \eqref{eq:big2}, where $\mathbf{\Delta}_{\mu}(\bm{\theta},\omega,\mathbf{u}) \triangleq \mathbf{H}(\bm{\theta} + \mu \mathbf{u},\omega) - \mathbf{H}(\bm{\theta} - \mu \mathbf{u},\omega)$. The proposed \textit{Zeroth-order Stochastic Gradient Ascent} (ZoSGA) method with WMMSE is provided in Algorithm \ref{algorithm1}.
\SetInd{0.25em}{0em}
\begin{algorithm}[t]
\caption{ZoSGA with WMMSE}
\label{algorithm1}
\SetAlgoNoLine
Initialize $\mathbf{W}^0$, $\bm{\theta}^0$, $\eta^0$, $\mu$. \\
\For{$t=0,\ldots,T$}{
    $\quad$Sample (i.i.d.) $(\mathbf{u}^{t+1},\omega^{t+1})$. \\
    $\quad$Find $\mathbf{W}^*(\bm{\theta}^t,\omega^{t+1})$ using WMMSE \cite{wmmseShi2011}. \\
    $\quad$Compute $\mathbf{G}_{\mu}(\bm{\theta}^t,\omega^{t+1},\mathbf{u}^{t+1})$ using \eqref{eq:big2} (by probing). \\
    $\quad$Choose $\eta^{t+1}$ and update     
     \begin{equation*}
        \bm{\theta}^{t+1} = \Pi_{\mathcal{K}} \left(\bm{\theta}^t + \eta^{t+1} \mathbf{G}_{\mu}(\bm{\theta}^t,\omega^{t+1},\mathbf{u}^{t+1}) \right),
    \end{equation*}
    \noindent $\ \ $ where $\Pi_{\mathcal{K}}(\cdot)$ is the Euclidean projection onto $\mathcal{K}$.}
    Return $\bm{\theta}^{t^*}$, where $\mathbb{P}(t^* = t) = \frac{\eta^n}{\sum_{i = 0}^T\eta^i}$.
\end{algorithm}

\begin{figure*}[t]
  \centering
  \begin{subfigure}[b]{0.33\textwidth}
  \centering
  \includegraphics[width=\textwidth, height=0.75692307692\textwidth]{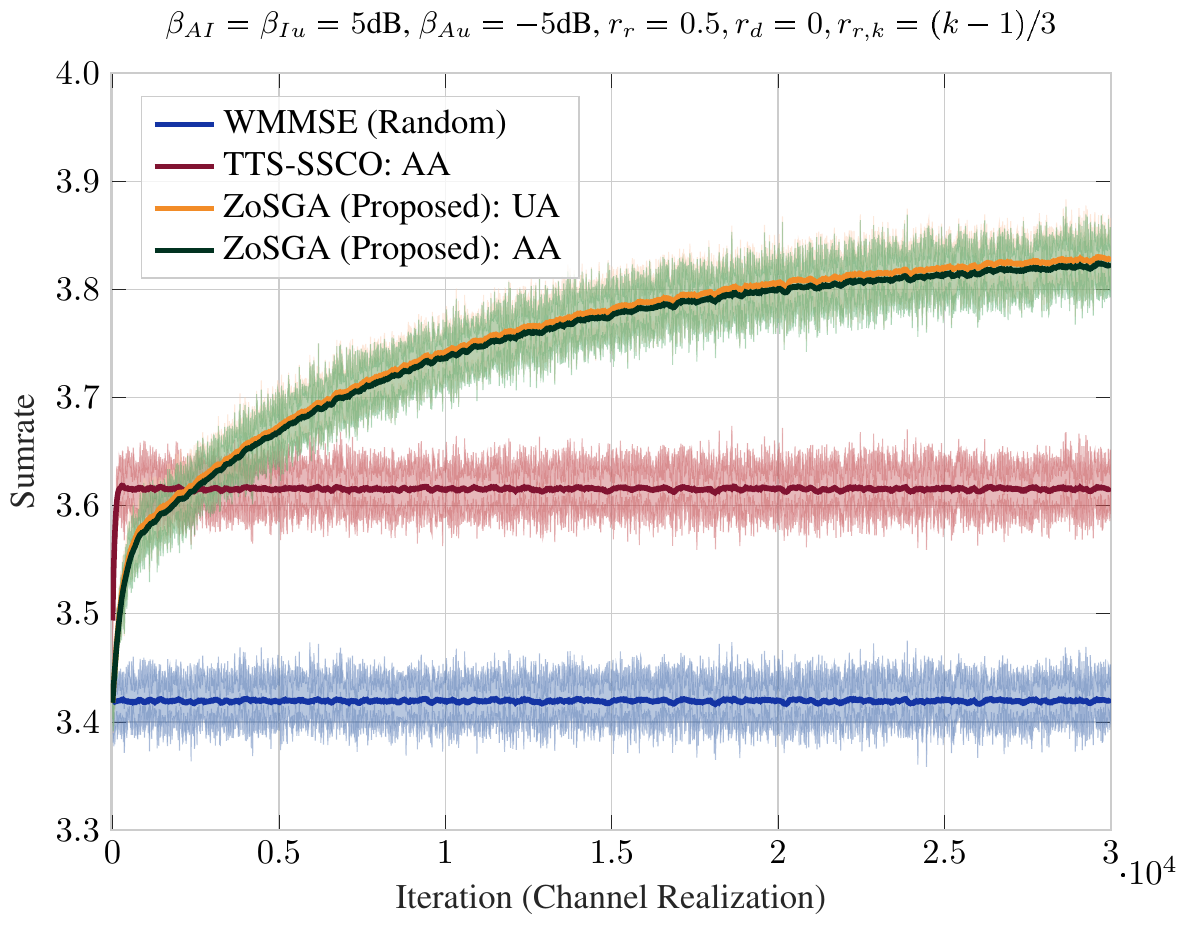}
  \caption{}
  \label{fig:converge1}
  \end{subfigure}
  \begin{subfigure}[b]{0.33\textwidth}
  \centering
  \includegraphics[width=\textwidth, height=0.75692307692\textwidth]{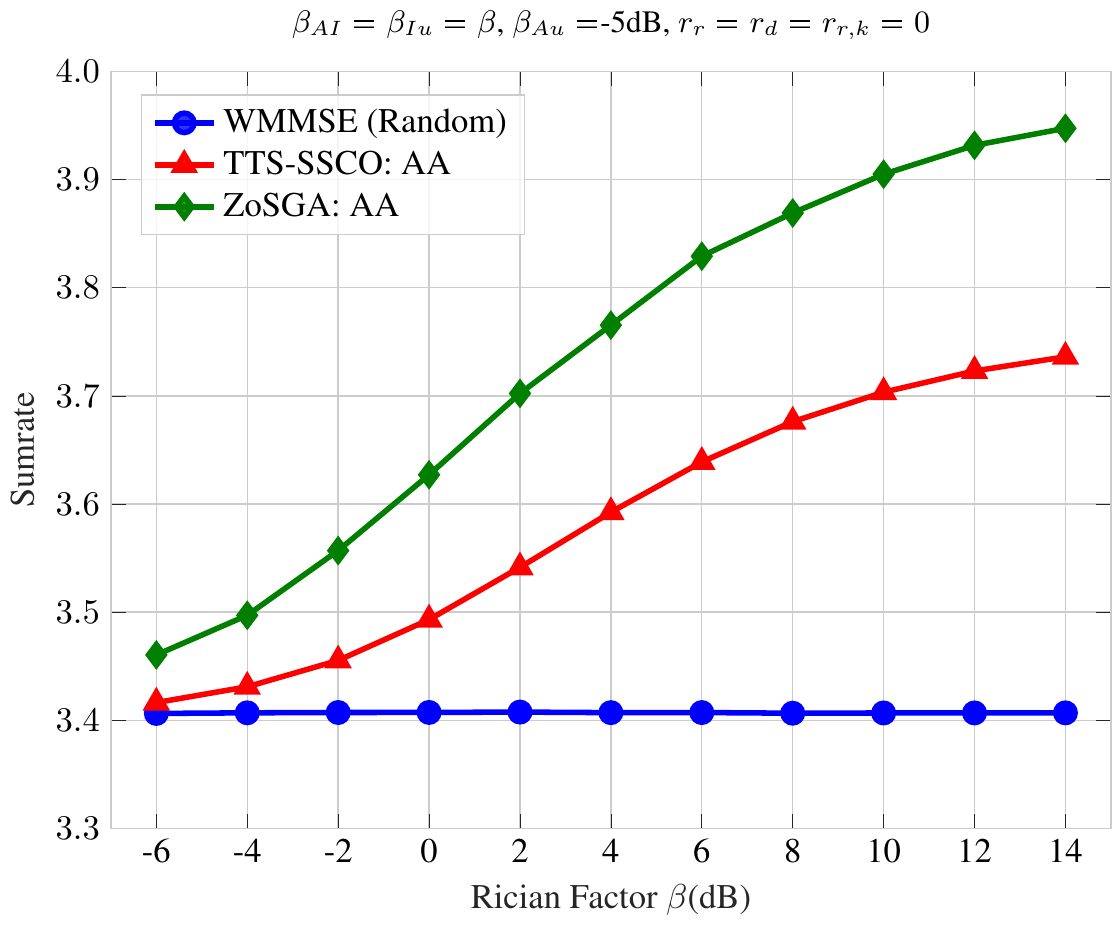}
  \caption{}
  \label{fig:rician}
  \end{subfigure}
  \begin{subfigure}[b]{0.33\textwidth}
  \centering
  \includegraphics[width=\textwidth, height=0.75692307692\textwidth]{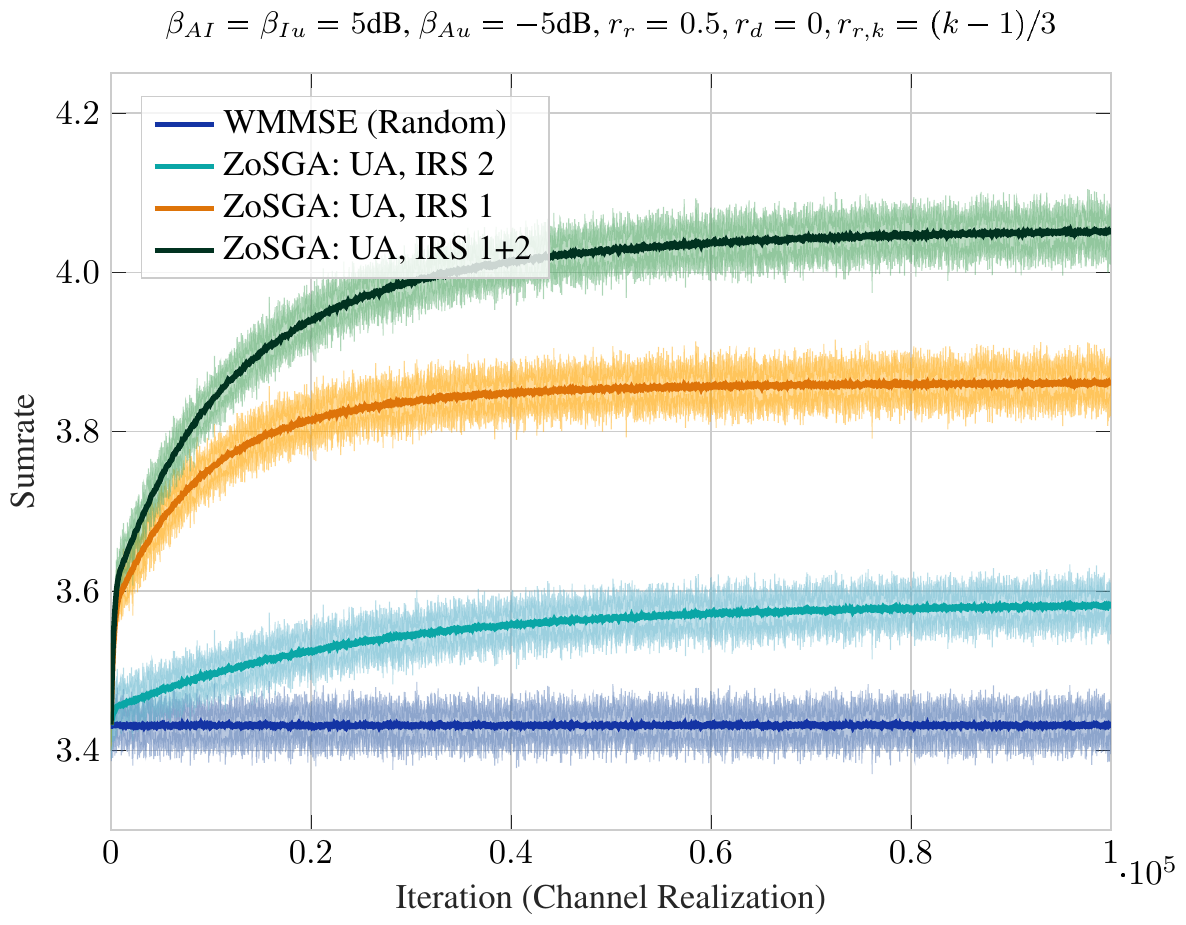}
  \caption{}
  \label{fig:converge2}
  \end{subfigure} 
\vspace{-10bp}
\caption{Average sumrates achieved by WMMSE \cite{wmmseShi2011} (with random IRS phase shifts), TTS-SSCO \cite{zhao2020tts}, and the proposed ZoSGA algorithm, with only IRS 1 present: (a) with fixed parameters $\beta_*$ and $r_*$, and (b) relative to Rician factors $\beta_{AI}= \beta_{Iu}$; (c) Average sumrates achieved by WMMSE and ZoSGA with IRS 1+2 present and fixed parameters $\beta_*$ and $r_*$. (AA: Adjustable Amplitude; UA: Unit Amplitude)}
\label{fig:main}
\vspace{-10bp}
\end{figure*}

\vspace{-10pt}
\section{Convergence analysis} \label{sec: conv anal}
\vspace{-8pt}
In this section we provide a brief account of the convergence properties of Algortithm \ref{algorithm1}. For convenience let $\varphi(\cdot) \triangleq -\mathbb{E}\{F_{\omega}^*(\cdot)\} + \delta_{\mathcal{K}}(\cdot)$, where $F_{\omega}^*(\cdot) \triangleq F\left(\mathbf{W}^*(\cdot,\omega),\mathbf{H}(\cdot,\omega)\right)$ is the (sample) objective of problem \eqref{eqn: outer problem} and $\delta_{\mathcal{K}}(\cdot) \in \{0,\infty\}$ is the indicator of the set $\mathcal{K}$. The following lemma is crucial for the analysis of Algorithm \ref{algorithm1}.
\begin{lemma} \label{lemma: weak convexity}
Let Assumptions \textnormal{\ref{assumption: inner problem solution}}, \textnormal{\ref{assumption: channel properties}} be in effect, and let $\widehat{\mathcal{K}} \supset \mathcal{K}$ be a convex compact set. Then, for a.e. $\omega$ there exists $\rho(\omega)>0$ such that $-F_{\omega}^*$ is $\rho(\omega)$-weakly convex on $\widehat{\mathcal{K}}$, i.e., $-F_{\omega}^*(\cdot) + (\rho(\omega)/2)\|\cdot\|^2$ is convex on $\widehat{\mathcal{K}}$.
\end{lemma}
\noindent Given some $\lambda > 0$, we define the \emph{Moreau envelope} of $\varphi$ as  
\[\varphi^{\lambda}(\bm{y})  \triangleq \min_{\bm{\theta}}\left\{\varphi(\bm{\theta}) + \frac{1}{2\lambda}\|\bm{y}-\bm{\theta}\|^2\right\}.\]
\noindent A near-stationary point for the Moreau envelope $\varphi^{\lambda}$ is close to a near-stationary solution of problem \eqref{eq:1}, if $-\mathbb{E}\{F_{\omega}^*(\cdot)\}$ is weakly convex (see \cite[Section 2.2]{SIAMOpt:Davis} and note, from  Lemma \ref{lemma: weak convexity}, that $-\mathbb{E}\{F_{\omega}^*(\cdot)\}$ is weakly convex if $\rho(\omega)$ is integrable). In what follows, we provide a convergence rate of Algorithm \ref{algorithm1} in terms of the gradient of the Moreau envelope.
\begin{theorem} \label{thm: convergence analysis}
Let Assumptions \textnormal{\ref{assumption: inner problem solution}}, \textnormal{\ref{assumption: channel properties}} be in effect and assume that $\mathcal{K}$ is convex and compact, $\rho = \mathbb{E}\{\rho(\omega)\} <\infty$, while the process $\{\omega^t\}_{t \in \mathbb{N}}$ is i.i.d.. Let $\{\bm{\theta}^t\}_{t = 0}^T$ be the sequence of iterates generated by Algorithm \textnormal{\ref{algorithm1}}, $\Delta_{\mathcal{K}} > 0$ be the diameter of $\mathcal{K}$, and set
\[ \eta^t = \sqrt{\frac{\Delta_{\varphi}}{4\rho B_F^2 L_{\mathbf{h},0}^2 (S^2 + 2S)(T+1)}},\qquad \textnormal{for all }t\geq 0,\]
\noindent for some $\Delta_{\varphi} \geq \varphi^{1/(2\rho)}(\bm{\theta}^0) - \min_{\bm{\theta}}  \varphi(\bm{\theta})$. Then, it is true that
\begin{equation*} 
    \begin{split}
     & \hspace{-4bp} \mathbb{E}\big\{\big\Vert \nabla \varphi^{1/(2\rho)}\big(\bm{\theta}^{t^*}\big)\big\Vert^2_2
     \big\} \\
     & \leq \ 8\Bigg(\sqrt{\frac{\Delta_{\varphi} \rho B_F^2 L_{\mathbf{h},0}^2 (S^2 + 2S)}{T+1}}
     + \mu\rho \Delta_{\mathcal{K}} B_F L_{\mathbf{h},1}\sqrt{S M K}\Bigg).
    \end{split}
\end{equation*}
\end{theorem}
\begin{remark}
Note that choosing 
$\mu = \mathcal{O}(1/
\sqrt{(M K T)})$
yields that 
$\mathbb{E}\big\{\big\Vert \nabla \varphi^{1/(2\rho)}\big(\bm{\theta}^{t^*}\big)\big\Vert_2
     \big\} \leq \epsilon$, 
after $\mathcal{O}(\sqrt{S}\epsilon^{-4})$ iterations.
\end{remark}
\vspace{-10pt}
\section{Simulations} \label{sec: Simulations}
\begin{figure}
  \centering
  \centerline{\includegraphics[width=2.6in]{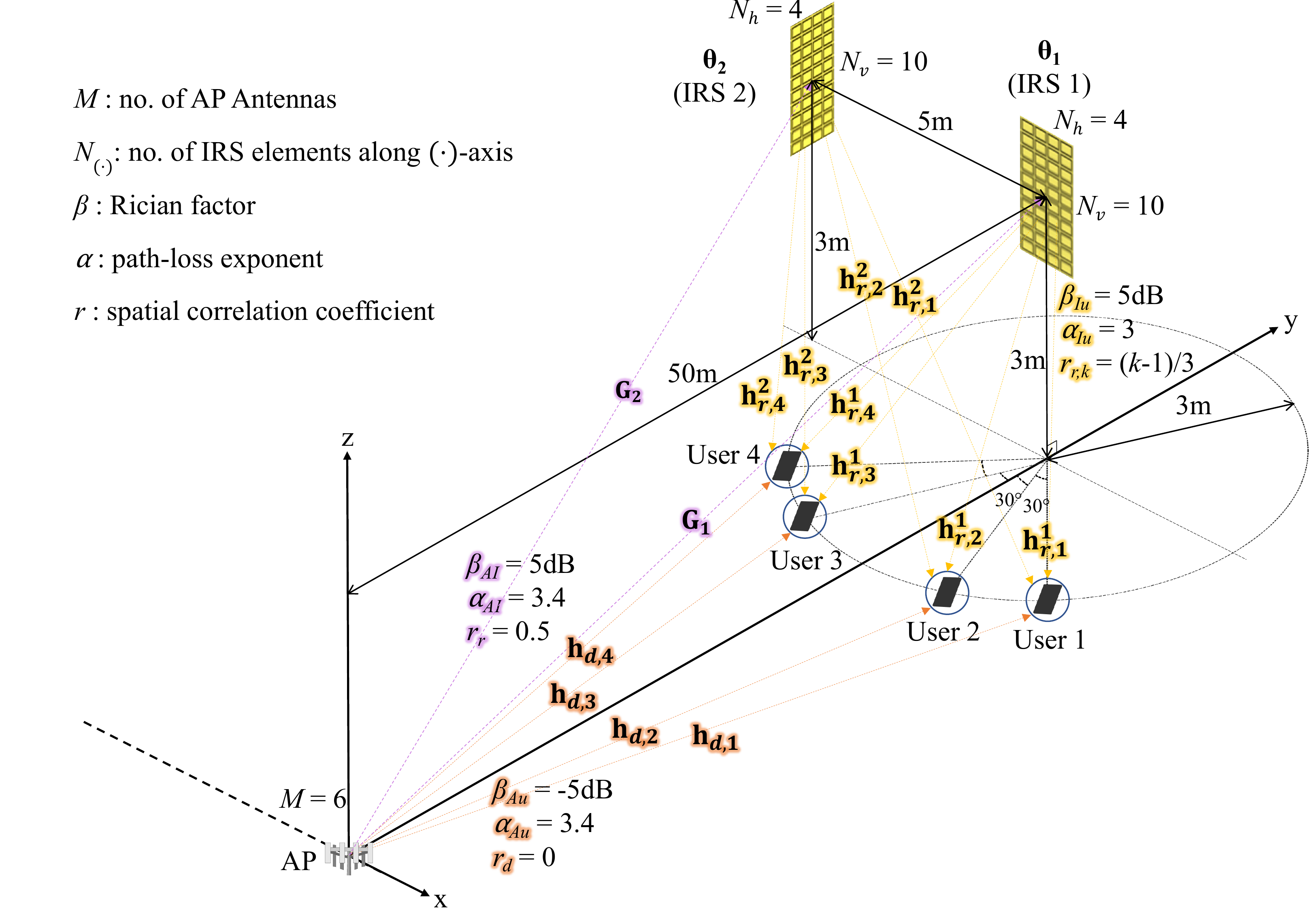}}
\vspace{6bp}
\caption{Experimental IRS-Aided Network Configuration.}
\label{fig:env_setup}
\end{figure}

\vspace{-4pt}
We consider an IRS-aided MISO downlink wireless network with Rician fading channels $\mathbf{G}_i \in \mathbb{C}^{N {\times} M}$, $\mathbf{h}^i_{r,k} \in \mathbb{C}^{N}$ and $\mathbf{h}_{d,k} \in \mathbb{C}^{M}$, respectively, as shown in Fig.\:\ref{fig:env_setup}, where $i \in \{1,2\}$. Following \cite{zhao2020tts}, these channels are defined for each user $k$ as 
\begin{align} 
\begin{aligned} \label{eq:rician_model}
    \mathbf{h}_{r,k}^{i} &{\triangleq} \sqrt{\beta_{Iu}{/}(1{+}\beta_{Iu})} \check{\mathbf{v}}^{i}_{r,k} {+} \sqrt{1{/}(1{+}\beta_{Iu})} \mathbf{\Phi}_{r,k}^{1/2}\mathbf{v}_{r,k}^{i},\\
    \mathbf{G}_{i} &{\triangleq} \sqrt{\beta_{AI}{/}(1{+}\beta_{AI})} \check{\mathbf{F}}^{i} {+} \sqrt{1{/}(1{+}\beta_{AI})} \mathbf{\Phi}_r^{1/2}\mathbf{F}^{i}\mathbf{\Phi}_d^{1/2},\\
    \mathbf{h}_{d,k} &{\triangleq} \sqrt{\beta_{Au}{/}(1{+}\beta_{Au})} \check{\mathbf{v}}_{d,k} {+} \sqrt{1{/}(1{+}\beta_{Au})} \mathbf{\Phi}_d^{1/2}\mathbf{v}_{d,k},\\
\end{aligned}
\end{align}
where the entries of $\mathbf{H}_{\text{S-CSI}} {\triangleq} \{\check{\mathbf{v}}^{i}_{r,k}{,} \check{\mathbf{F}}^{i}{,} \check{\mathbf{v}}_{d,k}\}$ and $\mathbf{H}_{\text{I-CSI}} {\triangleq}$ $\{\mathbf{v}^{i}_{r,k} {,} \mathbf{F}^{i} {,}$ $\mathbf{v}_{d,k}\}$ ($i \in \{1,2\}$) are all sampled independently from $\mathcal{CN}(\bf{0},\bf{I})$,
with the former being sampled only once per simulation, where \textit{a simulation} is defined as an ensemble containing a number of sequential i.i.d. realizations of "instantaneous" CSI $\mathbf{H}_{\text{I-CSI}}$ (given a fixed value for "statistical" CSI $\mathbf{H}_{\text{S-CSI}}$ for each simulation).

Each IRS has $N {=} N_{h} {\times} N_{v}$ phase-shift elements (see Fig.\:\ref{fig:env_setup}). Thus, $\bm{\theta}_i {=}  \begin{bmatrix} \bm{\phi}_i^\trans & \hspace{-6pt}\mathbf{A}_i^\trans \end{bmatrix}^\trans$, where $\bm{\phi}_i \in [-\pi,\pi]^{N}$ and $\mathbf{A}_i \in [0,1]^{N}$ are phases and amplitudes of the IRS elements, respectively, with $i \in \{1,2\}$. The scalars $\beta_{Iu}$, $\beta_{AI}$, and $\beta_{Au}$, are link-specific Rician factors, while $\mathbf{\Phi}_{r,k} \in \mathbb{C}^{N {\times} N}$, $\mathbf{\Phi}_{r} \in \mathbb{C}^{N {\times} N}$ and $\mathbf{\Phi}_d \in \mathbb{C}^{M {\times} M}$ are spatial correlation matrices following the exponentially decaying correlation model of \cite{zhao2020tts,zhao2021qos}, 
with corresponding correlation coefficients $r_{r,k}\in(0,1)$, $r_r\in(0,1)$ and $r_d\in(0,1)$.
Lastly, the distance-dependent path-loss for each link is modeled as $L = \sqrt{C_0{d}^{-\alpha}}$, where $d$ is distance (in meters), $\alpha$ is the corresponding path loss exponent, and $C_0 {=} $-30dBm.

Throughout the simulations we consider a power allocation of $P {=} $5dBm, a noise variance $\sigma^2 {=}$ -80dBm, and run WMMSE for 20 iterations to optimize the precoding vectors $\mathbf{W}$ for each instance of the received effective channels $\mathbf{h}_k(\bm{\theta},\omega)$, each of which is given as 
\begin{align} \label{eq: sim_model}
    \mathbf{h}_k(\bm{\theta},\mathbf{\omega}) =
    \sum_{i=1}^{2} \underbrace{\mathbf{G}^\hermtr_{i} \text{diag}(\mathbf{A}_i \circ e^{-j{\bm{\phi}}_i}) \mathbf{h}^{i}_{r,k}}_{\bm{\theta}_i\text{-reflected link}} +
    \underbrace{\mathbf{h}_{d,k}}_{\text{LoS link}},
\end{align}
for all $k \in \mathbb{N}^+_{4}$, where $\bm{\theta} {=} (\bm{\theta}_1, \bm{\theta}_2)$, $\omega=\{\mathbf{G}_1,\mathbf{G}_2, \mathbf{h}^{1}_{r,k}, \mathbf{h}^{2}_{r,k}, \mathbf{h}_{d,k},$ $k=1,\ldots,K\}$ and, for simplicity, we ignore IRS-to-IRS links. All presented results are averaged over 2000 unique simulations.

First, we compare the proposed ZoSGA against TTS-SSCO \cite{zhao2020tts} for a network with one IRS as TTS-SSCO requires an exact model to optimize the IRS phase-shift elements (see\:\cite[eq.\:(35)]{zhao2020tts}). We let $\mu{=}10^{-12}$, and choose separate step-sizes $\eta^0_{\phi} = 0.4$ and $\eta^0_{A} = 0.01$ for $\bm{\phi}$ and $\mathbf{A}$, respectively scaled by ${0.9972}^t$ for $t \in \mathbb{N}^+_{10^3}$, keeping them constant afterwards. The parameters for TTS-SSCO are provided in \cite[Section V]{zhao2020tts}. From Fig.\:\ref{fig:converge1}, we observe that ZoSGA substantially outperforms TTS-SSCO solely on the basis of effective CSI, while having \textit{no} access to the statistical model of the channel or the spatial configuration of the system; this is in sharp contrast to TTS-SSCO. While TTS-SSCO converges faster, it does so by internally sampling I-CSI and finding corresponding optimal precoding vectors (via WMMSE) 10 times per iteration. In Fig.\:\ref{fig:rician}, we see that, for spatially uncorrelated channels, the relative gain of ZoSGA in the achievable sumrate increases with respect to the Rician factors pertaining to $\bm{\theta}_1$-reflected links, i.e., as we move from I-CSI to S-CSI dominated channels. We also observe that WMMSE with a randomized IRS is insensitive to changes in Rician factor; this is expected.

Lastly, we use both IRSs and optimize their phase-shift elements using identical stepsizes for ZoSGA as stated above. Our results in Fig.\:\ref{fig:converge2} show that ZoSGA not only scales well to unknown system/channel models, but that it is also robust to the choices of $\eta_{\phi}$ and $\eta_{A}$. The results also demonstrate the better performance gains when optimizing $\bm{\theta}_1$ versus optimizing the more distant $\bm{\theta}_2$.
\vspace{-7pt}
\section{Conclusion}
\vspace{-7pt}
We introduced ZoSGA, a truly model-free method, to optimize fully-passive IRS phase-shift elements for a given QoS metric. We proved state-of-the-art convergence rates for ZoSGA under standard assumptions, and empirically demonstrated its efficiency on a benchmark IRS-aided network. In fact, ZoSGA outperforms the current SOTA by learning (near-)optimal passive IRS beamformers solely on the basis of conventional effective CSI, in the absence of channel models and spatial network configuration information.
\clearpage
\bibliographystyle{IEEEbib-abbrev}
\bibliography{zosga.bib}
\end{document}